# Observations and Analysis of Three Field RR Lyrae Stars Selected Using Single epoch SDSS Data

**W. Lee Powell, Jr.**
**Stephanie N. Jameson**
*University of Nebraska Kearney, Department of Physics, Bruner Hall of Science, 2401 11th Avenue, Kearney, NE 68849; address email correcpondence to W. L. Powell, Jr., powellwl@unk.edu*

**Nathan De Lee**
*Northern Kentucky University, Department of Physics and Geology, Natural Science Center 204H, Nunn Drive, Highland Heights, KY 41099; and Vanderbilt University, Department of Physics and Astronomy, Nashville, TN 37235*

**Ronald J. Wilhelm**
*University of Kentucky, Department of Physics and Astronomy, 177 Chemistry-Physics Building, 505 Rose Street, Lexington, KY 40506*

**Abstract**    We present the results of our Johnson B and V observations of three RR Lyrae candidate stars that we identified as likely variable stars using SDSS data. The stars were selected based upon a single epoch of photometry and spectroscopy. The stars were observed at McDonald Observatory to obtain full light curves. We present full light curves, measured periods, and amplitudes, as well as the results of our Fourier analysis of the light curves.

## 1. Introduction

In an effort to obtain as much useful information as possible out of large survey datasets, we derived a means of using a single epoch of photometry and spectroscopy data to identify RR Lyrae variable stars. Our method takes advantage of correlated changes in observed colors versus effective width of lines in the spectrum to find stars that have different surface temperatures at the two pulsation states. We found a large disparity between the (g–r) color and the strength of the Hydrogen Balmer lines when two observations were made at random phase, which we exploit to identify RR Lyrae pulsating variable stars. As a result of that effort we identified over 1,000 candidate RR Lyrae stars in the halo of the Milky Way ranging from around g=14 to g=20 using Sloan Digital Sky Survey data (York *et al.* 2000). This dataset is not complete since it will only identify stars that were at different points of their pulsation cycle when the photometry and spectroscopy observations were made (Wilhelm *et al.* 2007; Powell *et al.* 2010). Testing the method using the known variables identified in the SDSS Stripe 82 (Sesar *et al.* 2007) shows that our method is ~85% accurate in predicting variability. This accuracy has been confirmed by previous work by the authors (Powell *et al.* 2010). For that paper an assortment of candidate stars was observed sparely to confirm their variability. Since RR Lyrae stars have such large amplitudes they are easily identified even using sparse data. Our previous work combined with on-going work agrees well with the results determined using Stripe 82, with 15 stars confirmed as variable, three stars that appear to be variable but not RR Lyraes, and two stars that appear to not be variable, out of 20 stars observed between 2009 and 2014. Undergraduate students have made the vast majority of the photometry observations described in this paper. The three stars described in this paper appear to be previously unknown variables.

## 2. Observations and calibration

### 2.1. General observation information

All photometry observations described in this paper were made at McDonald Observatory using the 0.8-m telescope and the f/3.0 prime focus corrector (PFC). The PFC uses a Loral-Fairchild nitrogen-cooled 2048 × 2048 pixel CCD chip. The Johnson B and V filters were used for all observations. The readout time for this chip is quite long and the field of view is larger than needed so only the central portion of the chip was used. Using a 750 × 750 pixel region resulted in a field of view of around 15 arcminutes and a readout time of just over 70 seconds. This allowed us to achieve a faster cadence and more complete light curves. Dark current is negligible for this LN2-cooled instrument. Dome flat field frames and bias frames were taken nightly for all observing runs.

Table 1: Details of observations.

| Star | SDSS ID | R. A. (2000) h | Dec. (2000) ° | Dates | Observers |
|---|---|---|---|---|---|
| RR 444 | 4828-301-5-116-0447 | 266.368714 | +26.141512 | May 27–30, 2011 | Powell, Hans Amende, Caleb Bahr |
|  |  |  |  | July 5–8, 2011 | Powell, Hans Amende, Caleb Bahr |
| RR 143 | 4512-301-3-0275-0084 | 240.399352 | +23.467972 | June 6–12, 2013 | Powell, Stephanie Smith |
| RR 397 | 3705-301-2-0292-0054 | 245.745375 | +36.573442 | June 6–12, 2013 | Powell, Stephanie Smith |

Table 2. Comparison stars.

| Variable | APASS Comparison | R. A. (2000) h | Dec. (2000) ° | Number of Observations | APASS Magnitudes | | | | |
|---|---|---|---|---|---|---|---|---|---|
| | | | | | V | V error | B | B error | B–V |
| RR 143 | C1 | 240.40006 | 23.50305 | 4 | 15.65 | 0.00 | 16.38 | –0.06 | 0.73 |
| | C3 | 240.51872 | 23.41518 | 2 | 15.96 | 0.00 | 16.69 | –0.07 | 0.73 |
| | C5 | 240.36162 | 23.45008 | 4 | 15.20 | 0.04 | 15.99 | 0.20 | 0.79 |
| | C6 | 240.37839 | 23.43636 | 4 | 12.78 | 0.10 | 15.59 | 0.20 | 0.81 |
| | C7 | 240.44765 | 23.46511 | 4 | 15.50 | 0.02 | 15.91 | 0.12 | 0.86 |
| RR 397 | C1 | 245.71351 | 36.560951 | 2 | 15.39 | 0.01 | 15.82 | 0.09 | 0.42 |
| | C2 | 245.81440 | 36.558467 | 2 | 14.69 | 0.01 | 15.27 | 0.03 | 0.58 |
| | C5 | 245.77940 | 36.556515 | 2 | 14.74 | 0.01 | 15.41 | 0.01 | 0.67 |
| RR 444 | C1 | 266.40591 | 26.235944 | 2 | 14.72 | 0.07 | 15.29 | 0.06 | 0.57 |
| | C2 | 266.44629 | 26.141434 | 2 | 15.81 | 0.00 | 16.09 | –0.05 | 0.29 |
| | C4 | 266.33562 | 26.136753 | 2 | 15.65 | 0.03 | 16.20 | 0.10 | 0.54 |
| | C5 | 266.31716 | 26.162361 | 2 | 15.35 | 0.07 | 15.93 | 0.02 | 0.58 |

### 2.2. Observing runs

Results for three stars are presented in this paper. The full SDSS names, coordinates, and observing run dates for these stars are listed in Table 1. Star RR444 was observed for a total of seven nights, four in May 2011 and three in July 2011. The two stars RR143 and RR397 were both observed during our seven night June 2013 observing run.

### 2.3. Data processing

All data were processed using the CCDPROC routine in IRAF. A master bias frame was made by median combining the bias frames from each observing run. An ovserscan correction was also made. Flat field images from the McDonald 0.8-m require additional processing to remove a gradient introduced by the geometry of the telescope, screen, and lighting system. The IRAF routine IMSURFIT was used to remove the geometry-induced gradient. The corrected flat field images were median combined to create a master flat field image for each night. The process CCDPROC was used to bias- and flat-correct the science frames, and was also used to correct bad pixels in an automated fashion. The final processed images from each night were then registered and aligned to make the photometry process simpler.

### 2.4. Comparison stars and photometry measurements
#### 2.4.1. Comparison stars

All comparison stars used for this project were found in the AAVSO Photometric All-Sky Survey (APASS; Henden *et al.* 2014). The AAVSO APASS online search tool was used to search the area around each variable star, within the field of view of the images. The final comparison stars were chosen based on their proximity to each variable star, as well as their similarity in color. By choosing comparison stars that are very similar in color to the variable star, we minimized error induced by differential affects of the atmosphere on each star's light due to its color. The comparison stars used are detailed in Table 2. None of the comparison stars used exhibited any noticeable variability more than the scatter one would expect from the errors in the photometry.

#### 2.4.2. Photometry measurements with MIRA

Since the intention of this project, in part, was to introduce undergraduate students to photometry, the use of a commercial photometry package was useful to speed the learning process. MIRA Pro Ultimate Edition (Mirametrics 2010) provides a tool for doing calibrated differential photometry. While identifying stars in each frame, the tool allows the appropriate magnitude to be entered for each standard star. The software then calculates a plate solution that returns a calibrated magnitude for each star in the image. The argument is that since all stars in the image are recorded at the same airmass and time, and since the color of the standards is approximately the same as that of the variable, then the magnitudes returned should be at the very least approximately on the standard system. The advantage with this approach when working with undergraduate students is that the math is automatically performed by the software, and the graphical interface is very user-friendly. The obvious disadvantage is that it is not immediately obvious that the resulting magnitudes are truly standard. This approach was used to extract magnitudes for the standards and variable star for all of the science images.

#### 2.4.3. Determination of extinction coefficients

To test that it is indeed the case that the resulting magnitudes are truly standard, the photometry was performed using a traditional approach for one night per variable star. This will provide evidence both that the standard stars chosen are consistent in their behavior, and that the MIRA approach yields the same answers as are obtained by calculating the magnitudes from instrumental magnitudes and extinction coefficients calculated for each night.

The approach used to determine the extinction coefficients follows closely the approach advocated by Peter Stetson (Stetson 1992). A full solution would follow this form:

$$v = V + a_0 + a_1(B-V) + a_2(B-V)^2 + a_3(B-V)^3 \\ + a_4(X-1.25) + a_5(X-1.25)(B-V) + a_6 t \quad (1)$$

where v is the instrumental magnitude, V is the standard magnitude, B–V is the color index, X is the airmass, and t is the Universal Time of the observation. Rather than using just X, the equation uses $(X-1.25)$ to avoid ringing in the numerical solution. The coefficients, **A**, that result from solving the system of simultaneous equations for all standard star observations, can then be used to solve for the measured magnitude on the standard system from the measured instrumental magnitude and instrumental color. Care must be taken to determine the instrumental magnitude, which usually means solving an abbreviated version (Equation 2) of the above equation to find preliminary coefficients that can be used to get an improved estimate of the instrumental color index for input into the longer equation for use in determining the full set of coefficients.

Table 3. Extinction coefficients.

| Coefficient | RR 143 Night 1 V | B | RR 397 Night 7 V | B | RR 444 Night 2 V | B |
|---|---|---|---|---|---|---|
| $a_0$ | −21.55160 | −21.1404 | −21.6222 | −21.189 | −21.5376 | −21.0956 |
| $a_1$ | 0.45330 | 0.7428 | 0.1365 | 0.3615 | 0.195 | 0.3358 |
| $a_2$ | 0.02700 | 0.0223 | 0.0041 | −0.0046 | 0.0073 | 0.0149 |

$$v = V + a_0 + a_1 (X - 1.25) + a_2 t \quad (2)$$

For this paper, the abbreviated form was used exclusively since the color indices were all quite similar. Since the numerical sampling in B–V is small, the resulting coefficients are unreliable and prone to ringing. The approach to solving this problem for this paper thus involved measuring the instrumental magnitude for the variable star and the standard stars, and recording X and t for each observation. Two matrices were then formed from the observations. The first matrix, **M**, has one row for each observation with column entries of the form:

$$1 \quad (X - 1.25) \quad t \quad (3)$$

A second column matrix, **m**, was created with each row formed from the corresponding number that results from subtracting the published standard magnitude from the instrumental magnitude for each observation, that is (v–V). Using tools from linear algebra we can solve for the coefficients using these matrices using the following equation:

$$\mathbf{A} = (\mathbf{M}^T \cdot \mathbf{M})^{-1} \cdot \mathbf{M}^T \cdot \mathbf{m} \quad (4)$$

which takes inverse of the matrix that results from taking the transpose of **M** times itself, multiplying that times the transpose of **M**, then multiplying that times the column matrix **m**. The result is a column matrix **A** with coefficients $a_0$, $a_1$, and $a_2$, where $a_0$ is the zero point and $a_1$ and $a_2$ are the slope terms for airmass and time. These matrix operations can be complete in a number of ways. For this paper we used the commercial package MATLAB (MathWorks 2014). This process was used to determine the coefficients for the V filter for one night per variable star, and repeated to find the coefficients for the B filter. The coefficients that were found are detailed in Table 3. The resulting coefficients were used to calculate the magnitudes of the stars in both V and B for comparison with the magnitudes that resulted from the use of MIRA.

2.4.4. Comparing the photometry results

For each variable star, the results of the two methods of photometry were compared for one night. The photometry results were compared to the published magnitude, and to each other. An average difference in magnitude was calculated, as was the standard deviation in the magnitude difference. The typical estimated error in the photometry, calculated by MIRA during the aperture photometry process using the signal-to-noise ratio, ranges from an average of around 0.005 to 0.03. As a preface to examining the differences between the two methods we note that the standard magnitudes for the RR 143 field are larger than those for the other fields, particularly in the B filter where the estimated APASS magnitude error grows to around 0.1. This seems to be echoed in our results. Table 4 summarizes the various comparisons between the two approaches.

We calculated the magnitude using two methods to confirm that the McDonald 0.8-m telescope data were transforming to the standard system reliably, and to explore whether using MIRA was introducing any systematic errors. The answer to the first question is yes. The result of both methods closely matched the standard system to the level of the error present in our photometry and the APASS photometry. No systematic problems were evident in our data. The answer to the question of whether using MIRA to calibrate the magnitude introduces systematic errors is shown to be no. The standard deviation in the magnitude difference of the MIRA photometry and the APASS standard magnitude was found to be smaller for both filters for all three nights than what we found doing an extinction calculation. Considering the errors in the photometry and the errors provided for the APASS data, any differences between the two methods are indistinguishable. Particularly reassuring is the fact that the average difference in all cases is very much smaller than the errors in the photometry. Considering all of the results we assert that the MIRA standard star correction seems to give reliable results, at least for the case of standard stars that are chosen to be similar in color to the variable star. Since this occurred for all three stars we chose not to perform the longer calculation for every night. The results of the MIRA photometry were used for the light curve fitting.

Table 4. Comparison of magnitudes found using MIRA and the extinction method.

| Star | MIRA vs APASS | | Calculation vs APASS | | MIRA vs Calculation | |
|---|---|---|---|---|---|---|
| | Average Difference | Standard Deviation | Average Difference | Standard Deviation | Average Difference | Standard Deviation |
| RR 143V | 0.000002564 | 0.05146 | −0.00354653 | 0.053817 | −0.003549 | 0.01062 |
| RR 143B | −0.00000303 | 0.1102 | 0.0001996 | 0.11158 | 0.00002613 | 0.01546 |
| RR 397V | −0.0000029 | 0.0009005 | −0.00018 | 0.01371 | −0.00077 | 0.010201 |
| RR 397B | 0.000958 | 0.0280009 | 0.00179 | 0.03145 | −0.00145 | 0.012158 |
| RR 444V | −0.007921 | 0.025381 | −0.00008135 | 0.02556 | −0.00784 | 0.009358 |
| RR 444B | −0.0000016 | 0.034901 | 0.000221 | 0.03686 | −0.00022 | 0.011602 |

## 3. Light curves and Fourier analysis

### 3.1. Period finding

The photometric analysis begins by identifying the correct period for each of the RR Lyrae stars. Given the sampling rate of this data set, there are a number of programs which can do this. For this work, SUPERSMOOTHER (Reimann 1994) was used. SUPERSMOOTHER uses a variable-span linear smoother that calculates a short, medium, and long smooth and then uses the best fit to the data. The period is determined by whichever frequency gives the best sum of absolute residuals. SUPERSMOOTHER has the benefit of being entirely automatic, and does not make assumptions about the shape of the light curve. SUPERSMOOTHER generates a list of the 15 most probable periods in order of likelihood.

### 3.2. Spline fitting

In order to determine the amplitudes, epochs of maximum, and weighted mean magnitudes of our stars it is useful to have a model of the light curve that is complete at all phases and can use information for each epoch in the light curve. There are several ways this could be achieved (fitting Fourier series, using previously defined templates, and so on), but in this paper we have chosen to use a smoothed spline. Normal splines weight each data point equally. A smoothed spline combines a spline and a chi square test to allow the error of each point to be taken into account. For this paper we use the smoothed spline algorighm described in Chapter 11 of Pollock (1999). The smoothed spline follows the actual shape of the data better than a template fit, and it is less prone to ringing than a Fourier series fit. The light curve properties for each star are given in Table 5.

### 3.3. Detemining metallicity

Due to the variable nature of RR Lyrae stars it is often difficult to get spectroscopic metallicities. As a result, a number of photometric methods have been developed to achieve such a measurement. Since the pulsation in RR Lyrae stars is driven by opacity changes in the atmosphere (the κ-mechanism (King and Cox 1968)), it is not surprising that metallicity would have an impact on the light curve shape. There are two techniques that take advantage of this fact, the $\Phi_{31}$-Period-metallicity (Jurcsik and Kovacs 1996) and Period-Amplitude-Metallicity (Sandage 2004) relations. Both are primarily used with type RRab stars, which have more variation in their light curve shape. Although progress has been made on the Fourier method with RRc stars in (Morgan *et al.* 2006), there are theoretical reasons to believe that the Period-Amplitude-Metallicity relation may not be as useful for RRc stars (Bono *et al.* 2007). For that reason, in this work both of these methods will be restricted to the RRab star.

### 3.4. Fourier metallicities

Fitting a Fourier sine series to a light curve provides a way to quantify the shape of the curve. In the Fourier sine series equation (Equation 5), $A_i$ is the amplitude of each order, $\varphi_i$ is the frequency offset, t is the HJD, and ω is the frequency. Once the sine series is fit to data, the adopted convention is to create a parameter $R_{ij} = A_i / A_j$ to relate the amplitudes and to relate the phases. For this work we fit an 11th-order sine series. We used the $\Phi_{31}$-period-metallicity relationship (Jurcsik and Kovacs (1996), shown in Equation 6) to determine the metallicity from the measured $\Phi_{31}$ and period.

$$V = A_0 + \sum_{i=1}^{n} A_i \sin [i\omega (t - t_0) + \varphi_i] \quad (5)$$

$$J_{[Fe/H]} = -5.038 - 5.39P + 1.345\, \Phi_{31} \quad (6)$$

In order to use Fourier sine fitting, a light curve must have relatively complete phase coverage, or else ringing will be induced in the fit. For those light curves that are too incomplete to use Fourier sine fitting we can use an amplitude-period-metallicity relationship. Sandage (2004) found a relationship among the amplitude in the Johnson V filter, the period, and the metallicity of an RRab star. His relationship is shown in equation (7).

$$[Fe/H] = -1.453\,(\pm 0.027)\, A_v - 7.990\,(\pm 0.091) \log P - 2.145 \pm 0.025 \quad (7)$$

In general the period-amplitude relation for determining metallicity has much more scatter in it relative to the Fourier method. This only makes sense given that it uses amplitude, which is much less sensitive to light curve shape than $\Phi_{31}$ (Bono *et al.* 2007). On the other hand, amplitude is a much easier quantity to measure and can be used on relatively incomplete light curves. Figure 7 shows RR 143 with the Fourier-fit to the V data. Table 5 details the parameters derived from our analysis of the light curves. The table lists for both V and B the amplitude and minimum magnitude of the light curve, the average magnitude derived from a magnitude-weighted and intensity-weighted average, the period of pulsation, the template that resulted in the best fit, the HJD of maximum magnitude, and the derived metallicities.

Table 5. Light curve derived parameters.

| Variable | V Amplitude | B Amplitude | Minimum V Mag. | Minimum B Mag. | V Mag. Weighted Average Mag. | V Intensity Weighted Average Mag. | B Mag. Weighted Average Mag. | B Intensity Weighted Average Mag. |
|---|---|---|---|---|---|---|---|---|
| RR 143 | 1.30 | 1.57 | 15.73 | 16.20 | 15.32 | 15.25 | 15.72 | 15.60 |
| RR 397 | 0.48 | 0.61 | 15.58 | 15.89 | 15.34 | 15.33 | 15.58 | 15.56 |
| RR 444 | 0.51 | 0.66 | 16.20 | 16.48 | 15.95 | 15.93 | 16.16 | 16.13 |

| Variable | Period V | Period B | HJD Max V | HJD Max B | [Fe/H]_J | [Fe/H]_S |
|---|---|---|---|---|---|---|
| RR 143 | 0.45543213 | 0.45543213 | 2456454.357537 | 2456454.35952526 | –1.46 | –1.32 |
| RR 397 | 0.36811239 | 0.37180909 | 2456456.889431 | 2456456.89763209 | NA | NA |
| RR 444 | 0.34790353 | 0.34790809 | 2455712.855586 | 2455712.850264 | NA | NA |

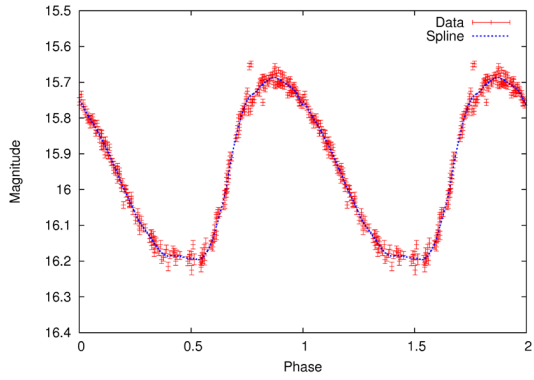

Figure 1. RR 444 Smoothed Spline-fit Lightcurve in V.

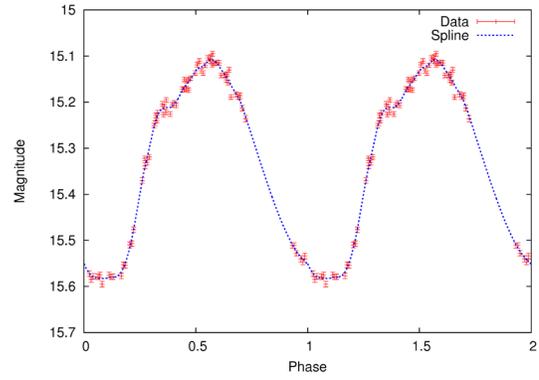

Figure 5. RR 397 Smoothed Spline-fit light curve in V.

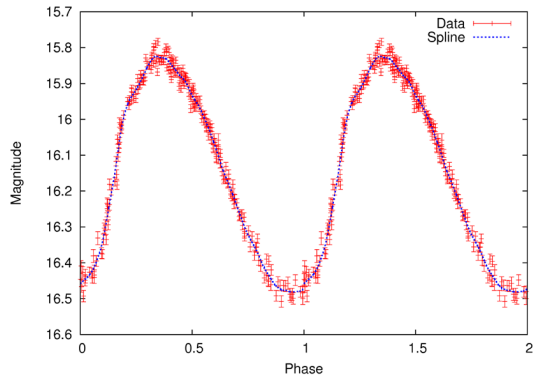

Figure 2. RR 444 Smoothed Spline-fit light curve in B.

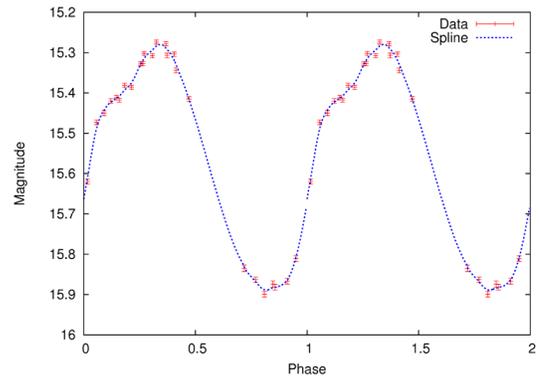

Figure 6. RR 397 Smoothed Spline-fit light curve in B.

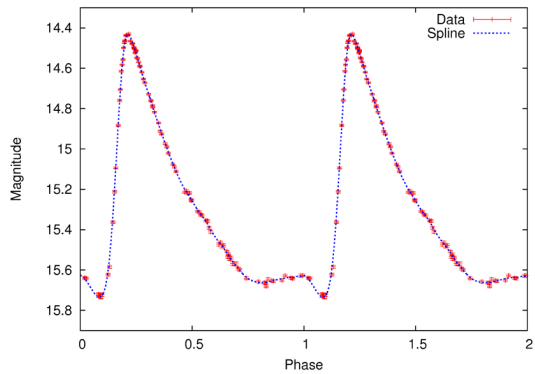

Figure 3. RR 143 Smoothed Spline-fit light curve in V.

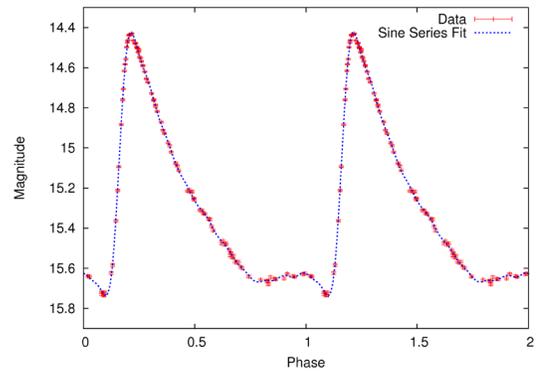

Figure 7: RR 143 Fourier-fit light curve in V.

### 4. Results

Figures 1 through 6 present results for three RR Lyrae stars. These figures show the light curves of each star for both B and V, including the observational data with error bars and the smoothed spline-fit results. RR 444 and RR 397 are RRc's. RR 143 is a RRab. The parameters derived from our analysis are presented in Table 5. The amplitudes for all three stars are typical for their types. As previously mentioned, only the metallicity derived from the Fourier analysis of the V data for RRab stars has meaning. That measurement yields for RR 143 a [Fe / H] value of −1.49 ± 0.054 derived from the Jurcsik and Kovacs method, and −1.37 ± 0.26 using Sandage's method. Both of these agree well with what might be expected for a halo field star. For details on this calculation and the estimated errors, see Jurcsik and Kovacs (1996) and Sandage (2004).

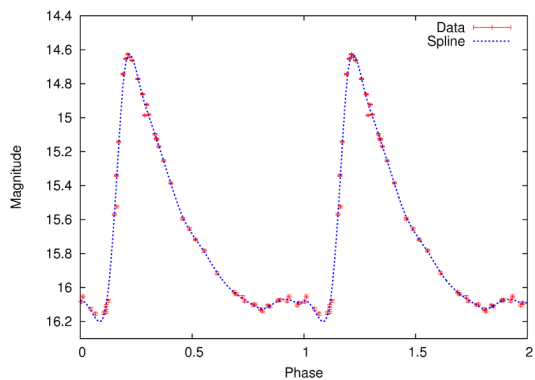

Figure 4. RR 143 Smoothed Spline-fit light curve in B.

The results presented are for three previously unknown variables. It is important to note that these were identified from the SDSS using only a single epoch of photometry and a single epoch of spectroscopy. This again serves as confirmation of our method of identifying RR Lyrae stars with single observations in large surveys.

This served as an excellent project to introduce undergraduate students to observational astronomy in general, and photometry specifically.

## 5. Future and on-going work

We are continuing to observe a large sample (N > 1,000) of suspected variables. We ultimately plan to continue working to confirm variability for much of the set. We have at this point confirmed variability for several more stars and are working on obtaining complete light curves for these confirmed variables. The area surrounding our RR Lyrae stars that is within the field of view of the images is being searched for additional new variable stars. All of this work is very accessible to undergraduate students and they will continue to be heavily involved in this project. AAVSO members interested in observing a sub-sample of these stars to help with either the confirmation of variability for the candidate stars or with the effort to obtain complete light curves may contact Dr. Powell to request more information on how to help.

## 6. Acknowledgements

The authors would like to thank McDonald Observatory for generous allocations of observing time to Dr. Powell and his students on the 0.8-m telescope. Dr. Powell would like to thank Nebraska EPSCoR for support of his student, Mrs. Jameson, in her research with a Summer Research at Undergraduate Institutions grant. Dr. Powell would also like to thank two undergraduate students that worked on the early stages of this project: Caleb Bahr and Hans Amende.

IRAF is distributed by the National Optical Astronomy Observatory, which is operated by the Association of Universities for Research in Astronomy (AURA) under cooperative agreement with the National Science Foundation.

This research was made possible through the use of the AAVSO Photometric All-Sky Survey (APASS), funded by the Robert Martin Ayers Sciences Fund.

This paper includes data taken at The McDonald Observatory of the University of Texas at Austin.

This paper was supported in part by NASA through the American Astronomical Society's Small Research Grant Program.